\documentclass[aps,prd,superscriptaddress,nofootinbib,nobibnotes,groupedaddress]{revtex4}

\usepackage{setspace}
\usepackage{color}
\usepackage{amsmath}
\usepackage{amsfonts}
\usepackage{verbatim}
\usepackage{amssymb}
\usepackage{graphicx,bm}
\usepackage{amsmath}
\usepackage{amssymb}
\usepackage{graphicx}
\usepackage[caption=false]{subfig}

\providecommand{\U}[1]{\protect\rule{.1in}{.1in}}

\newcommand{\be}{\begin{equation}}
\newcommand{\ee}{\end{equation}}

\newcommand{\mincir}{\raise
-3.truept\hbox{\rlap{\hbox{$\sim$}}\raise4.truept\hbox{$<$}\ }}
\newcommand{\magcir}{\raise
-3.truept\hbox{\rlap{\hbox{$\sim$}}\raise4.truept\hbox{$>$}\ }}

\begin{document}

\title{Cosmological phase-space analysis of $f(G)$-theories of gravity}

\author{Giannis Papagiannopoulos}
\email{yiannis.papayiannopoulos@gmail.com}
\affiliation{Department of Physics, National \& Kapodistrian University of Athens,
Zografou Campus GR 157 73, Athens, Greece}

\author{Orlando Luongo}
\email{orlando.luongo@unicam.it}
\affiliation{School of Science and Technology, University of Camerino, Via Madonna delle
Carceri, Camerino, 62032, Italy.}
\affiliation{Department of Nanoscale Science and Engineering, University at Albany-SUNY, Albany, New York 12222, USA.}
\affiliation{INAF - Osservatorio Astronomico di Brera, Milano, Italy.}
\affiliation{Istituto Nazionale di Fisica Nucleare (INFN), Sezione di Perugia, Perugia,
06123, Italy.}
\affiliation{Al-Farabi Kazakh National University, Al-Farabi av. 71, 050040 Almaty,
Kazakhstan.}

\author{Genly Leon}
\email{genly.leon@ucn.cl}
\affiliation{Departamento de Matem\'{a}ticas, Universidad Cat\'{o}lica del Norte, Avda.
Angamos 0610, Casilla 1280 Antofagasta, Chile}
\affiliation{Department of Mathematics, Faculty of Applied
Sciences, Durban University of Technology, Durban 4000, South Africa}

\author{Andronikos Paliathanasis}
\email{anpaliat@phys.uoa.gr}
\affiliation{Departamento de Matem\'{a}ticas, Universidad Cat\'{o}lica del Norte, Avda.
Angamos 0610, Casilla 1280 Antofagasta, Chile}
\affiliation{Department of Mathematics, Faculty of Applied
Sciences, Durban University of Technology, Durban 4000, South Africa}
\affiliation{School for Data Science and Computational Thinking and Department of
Mathematical Sciences, Stellenbosch University, Stellenbosch, 7602, South
Africa}

\begin{abstract}
The impact of topological terms that modify the Hilbert-Einstein action is
here explored by virtue of a further $f(G)$ contribution. In particular, we
investigate the phase-space stability and critical points of an equivalent
scalar field representation that makes use of a massive field, whose
potential is function of the topological correction. To do so, we introduce to the gravitational Action Integral a Lagrange multiplier and model the modified Friedmann
equations by virtue of new non-dimensional variables. We single out
dimensionless variables that permit \emph{a priori} the Hubble rate change of sign,
enabling regions in which the Hubble parameter either vanishes or becomes
negative. In this respect, we thus analyze the various possibilities
associated with a universe characterized by such topological contributions
and find the attractors, saddle points and regions of stability, in general.
The overall analysis is carried out considering the exponential potential
first and then shifting to more complicated cases, where the underlying
variables do not simplify. We compute the eigenvalues of our coupled
differential equations and, accordingly, the stability of the system, in
both a spatially-flat and non-flat universe. Quite remarkably, regardless of
the spatial curvature, we show that a stable de Sitter-like phase that can
model current time appears only a small fraction of the entire phase-space,
suggesting that the model under exam is unlikely in describing the whole
universe dynamics, i.e., the topological terms appear disfavored in framing
the entire evolution of the universe.
\end{abstract}

\maketitle

\section{Introduction}

Challenging General Relativity has acquired much more attention in view of
the most recent cosmological observations that certify the existence of a
dark sector \cite%
{2006IJMPD..15.1753C,Gruber:2013wua,delaCruz-Dombriz:2016bqh,Sahni:2006pa,Dunsby:2016lkw,Luongo:2025iqq},
fully-unknown, albeit necessary for both clustering and accelerating the
universe \cite{Bamba:2012cp,2001IJMPD..10..213C,2003PhRvL..90i1301L}.
Accordingly, conceptual limitations or experimental drawbacks are becoming
gradually more explored \cite{Simon2005}, especially in strong gravitational
fields \cite{2020A&A...641A.174L,2021MNRAS.503.4581L}, where gravity breaks down \cite{Kiefer:2023bld}  or a non-trivial mechanism of vacuum energy production is expected \cite{Martin:2012bt} leading to the cosmological constant problem \cite{Luongo:2018lgy,Belfiglio:2022qai,Luongo:2023aaq} and so on.

The advantage of extending the Hilbert-Einstein's action lies on the fact
that Einstein's theory is not renormalizable \cite {2003PhR...380..235P,1992ARA&A..30..499C} and, in turn, cannot be
conventionally quantized. The cosmological constant problem, for example,
remains an open challenge of our current understanding \cite%
{weinberg1989cosmological,Luongo:2018lgy,Capozziello:2025bsm} and limits the possibility to
characterize the cosmological constant today, as experimentally found \cite%
{2024A&A...690A..40L, Khadka:2021vqa, Luongo:2021nqh,Dunsby:2015ers,Vilardi:2024cwq,Bamba:2012cp,Capozziello:2021xjw}.

To face the problem of renormalization at one loop requires
the Einstein-Hilbert action to be supplemented with additional higher-order
curvature terms that, mostly turn out to be not unitary.
Analogously, when quantum corrections, for example under the form of string
modifications, are included, the effective low-energy gravitational action
naturally incorporates higher-order curvature invariants, leading to
experimental signature in extremely strong gravity regimes such the
cosmological Planck scale, black hole singularities, etc.

However, signatures of this picture may be found even at the level of
infrared scales, providing evidence in favor of modifying Einstein's
gravity at all scales.

A natural extension of Einstein's gravity may lead to $f(R)$ theories of
gravity \cite{Sotiriou:2008rp,Nojiri:2008nk,Nojiri:2008nt}, constructed as
analytic functions of the Ricci scalar $R$, whose prototype is offered by a
second order term, $\propto R^{2}$, named scalaron~\cite{Starobinsky:1980te}%
. Viable $f(R)$ models, fulfilling both cosmological and local gravity
constraints, have not really been found so far, and, in general, the matter sector
is mainly coupled with such theories, limiting the consistency with
observations.

Going through these paradigms, in view of the large number of proposals, one
can find theories with a Lagrangian density being a combination of terms~%
\cite{Barrow:1988xh,Barrow:1991hg,Bueno:2016dol} $R_{\mu \nu
}R^{\mu \nu }$ and $R^{\mu \nu \rho \sigma }R_{\mu \nu \rho \sigma }$,
namely superpositions of the Ricci and Riemann tensors, respectively, that are jeopardized by unphysical spin-2 ghost instabilities. We mention here, for the sake of completeness, other approaches such as torsion \cite{Nesseris:2013jea,Geng:2011aj,Paliathanasis:2016vsw,Duchaniya:2023aeu,DAgostino:2018ngy} or non-metricity \cite{Heisenberg:2023lru,Koussour:2023rly,Solanki:2022ccf,Anagnostopoulos:2021ydo} that can, as well, provide hints on how to facethe aforementioned issue, see e.g. \cite{BeltranJimenez:2019esp}. 

In this respect we single out the study of Gauss-Bonnet (GB) terms \cite%
{Padmanabhan:2013xyr,Dotti:2007az,Charmousis:2002rc,Pozdeeva:2019agu},
defined as, $G\equiv R^{2}-4R^{\mu \nu }R_{\mu \nu }+R_{\mu \nu \alpha \beta
}R^{\mu \nu \alpha \beta }$, that can significantly improve the above picture. {The GB scalar is the first topological scalar provided within the Lovelock theory of gravity \cite{Lovelock:1971yv,Lovelock:1972vz}. The latter theory is the second-order alternative to General Relativity gravitational theory in higher-dimensional geometries. However, in the case of a four-dimensional manifold Lovelock's topological scalars are reduced topopological boundary terms, and general relativity is recovered. Thus, in order to understand the effects of these scalars in low-dimensional geometries, nonlinear terms are introduced in the gravitational action integral.} 

Over the past decades, various studies have explored this topic, generally revealing that satisfying local gravity constraints \cite{DeFelice:2009aj,Yousaf:2019upm,Bhatti:2020det} remains challenging when the GB term is responsible for dark energy. However, this appears feasible for certain plausible $f(G)$ Lagrangians \cite{Nojiri:2005jg} that yield a cosmological constant \cite{Li:2007jm,Nojiri:2021mxf,Nojiri:2024nlx,Myrzakulov:2024sne,Lohakare:2024ize,Bamba:2017cjr,Nojiri:2010wj,Nojiri:2017ncd,Nojiri:2018ouv,Vernov:2021hxo}.

Although a definitive conclusion on the validity of extended gravity, and specifically $f(G)$ models, has yet to be reached, investigating their stability and properties remains crucial for assessing their potential role in cosmology.

Motivated by the above considerations, we here focus on $f(G)$ theories of
gravity and analyze the corresponding stability and phase-space dynamics. To
do so, we limit our study to a FLRW background, reconsidering $f(G)$
theories as scalar field-equivalent representation, modifying the
corresponding Lagrangian by including a Lagrangian multiplier. After
computing the modified Friedmann equations, we formulate the cosmological
dynamics by virtue of appropriate dynamical variables, rewritten in order to
account for all possible behaviors of the universe, namely enabling the
Hubble parameter to change sign, as a consequence of the dynamical
behaviour. The corresponding autonomous system of dynamical equations has
been obtained and the computation of eigenvalues has been found. Analyzing
then the phase-space we obtain the existence of attractors and explore the
stability of each point under examination. Following that the stability at asymptotic
regime also has been found. To do so, we employ a particular version of power-law
potential and compare our findings with previous works. As a matter of
fact, we emphasize that the so-constructed model converges to a de Sitter
phase

The paper is organized as follows. In Sect. \ref{sezione2}, we introduce the
concepts related to $f(G)$ cosmology. The equivalent scalar field
description is also reported in detail. The dynamical consequences, adopting
the modified Friedmann equations, are reported in Sect. \ref{sezione3}.
There, we also baptize our dynamical variables and set the system of
differential equations, used throughout our work. The consequences on our
observable universe are summarized in Sect. \ref{sezione4}. There, we report
the cases of spatially-flat and non-flat universe and the overall analysis.
Finally, the scalar field potentials and their impact under the form of
power-law contribution is explored in Sect. \ref{sezione5}, whereas our
final outlooks and perspectives are summarized in Sect. \ref{sezione6}.

\section{$f\left( G\right) $-Cosmology}\label{sezione2}

At large scales, the maximally symmetric spacetime describing the universe
is the four-dimensional isotropic and homogeneous
Friedmann-Lemaitre-Robertson-Walker geometry, 
\begin{equation}
ds^{2}=-N^{2}\left( t\right) dt^{2}+a^{2}\left( t\right) \left[ \frac{dr^{2}%
}{1-kr^{2}}+r^{2}\left( d\theta^{2}+\sin^{2}\theta d\varphi ^{2}\right) %
\right] ,  \label{cc.01}
\end{equation}
with $a\left( t\right) $ representing the cosmic scale factor, $N\left(
t\right) $ the lapse function and $k$ denotes the spatial curvature of the
associated three-dimensional hypersurface.

The latter indicates the universe topology, namely for $k=0$, the universe
is spatially flat, while for $k=-1$, the hypersurface has negative
curvature, turning out to be hyperbolic and, finally, $k=1$ leads to an
hypersurface represented by a closed sphere.

For the gravitational theory, we assume the modified four-dimensional
Gauss-Bonnet theory with the action, $S[R,G]=\int d^{4}x\sqrt{-g}\,\mathcal{L%
}[R,G]$, given by \cite{Li:2007jm} 
\begin{equation}
S[R,G]=\int d^{4}x\sqrt{-g}\left(\frac{R}{2}+f\left( G\right) \right) ,  \label{cc.02}
\end{equation}%
where $f\left( G\right) $ is an arbitrary function of the Gauss-Bonnet
scalar, $G$, and $R$ is the Ricci scalar, providing standard Einstein's
gravity.

As stated in the introduction, in a four-dimensional manifold, the
Gauss-Bonnet scalar is a topological invariant, i.e., $G$ represents a
boundary term, while considering any linear $f\left( G\right) $ function, 
\emph{de facto} lets Eq. \eqref{cc.02} turn into the equivalent
Hilbert-Einstein action.

Accordingly, we hereafter focus on non-linear $f\left( G\right) $
contributions, with the aim of exploring its effects over the cosmic
expansion history.

\subsection{Scalar field equivalent description}

Remarkably, introducing a Lagrange multiplier would easily provide an
equivalent description in terms of scalar fields. Accordingly, we can show
that the geometrodynamical degrees of freedom provided by the nonlinear
function $f\left( G\right) $ can be mimicked with a scalar field\footnote{%
A scalar field is commonly associated with interactions and particles.
It is always possible, however, to attribute dynamical properties of a given
geometrical model to it, with the great advantage of handling a
representation whose stability is easier to compute.}.

Specifically, the gravitational model in Eq. (\ref{cc.02}) is equivalent to
the Einstein-Gauss-Bonnet-Scalar field theory, where the scalar field is
coupled to $G$.

Motivated by this recipe, we employ the Lagrange multiplier, $\lambda$, and
write the Eq. (\ref{cc.02}) as 
\begin{equation}
S_{f\left( G\right) }=\int d^{4}x\sqrt{-g}\left( \frac{R}{2}+f\left( G\right)
-\lambda\left( G-\mathcal{G}\right) \right) ,  \label{cc.03}
\end{equation}
where, solving the equations of motion, leads to the constraint $G=\mathcal{G%
}$, that will be essential for our computation later on.

By varying the Lagrangian with respect to the Gauss-Bonnet term gives

\begin{equation}
\lambda=f_{,G}\left( G\right),
\end{equation}

that permits to reformulate the Lagrangian as

\begin{equation}  \label{cc.04}
\mathcal{L}= \frac{R}{2}+\left( f\left( G\right) -Gf_{,G}\left( G\right) \right)
+f_{,G}\left( G\right) \mathcal{G}.
\end{equation}

Conveniently, ensuring that $f(G)$ and $f_{,G}(G)$ are distinct variables, we
can introduce the scalar field representation by

\begin{subequations}
\begin{align}
& \phi =f_{,G}\left( G\right) , \\
& V\left( \phi \right) =\left( f\left( G\right) -Gf_{,G}\left( G\right)
\right) ,
\end{align}%
ending up with the equivalent Einstein-Gauss-Bonnet scalar field picture
\end{subequations}
\begin{equation}
S_{f\left( G\right) }=\int d^{4}x\sqrt{-g}\left( \frac{R}{2}+\phi G+V\left(
\phi \right) \right) .  \label{cc.05}
\end{equation}%
This is the Action Integral for the Einstein-Gauss-Bonnet Scalar field
gravitational model \cite%
{Fomin:2018typ,Odintsov:2020xji,Paliathanasis:2024gwp,Millano:2024vju,TerenteDiaz:2023iqk,Nojiri:2023jtf,Hussain:2024yee}
without a kinetic term.

Afterwards, from the line element in Eq. (\ref{cc.03}), the Ricci, $R$, and
the Gauss-Bonnet, $G$, scalars are 
\begin{subequations}
\begin{align}
&R=6\left( \frac{1}{N}\dot{H}+2H^{2}+\frac{k}{a^{2}}\right),  \label{cc.06}
\\
&G=24\left( \frac{1}{N}\dot{H}+H^{2}\right) \left( H^{2}+\frac{k}{a^{2}}%
\right),  \label{cc.07}
\end{align}

where we consider the Hubble function to be, $H\equiv\frac{1}{N}\frac{\dot{a}}{a}$.

Hence, plugging the Ricci and Gauss-Bonnet scalars into Eq. (\ref{cc.05})
and, thus, integrating by parts, we obtain the following point-like
Lagrangian, 
\end{subequations}
\begin{equation}
\mathcal{L}\left( N,a,\dot{a},\phi,\dot{\phi}\right) =\frac{3}{N}a\dot {a}%
^{2}-\frac{8}{N^{3}}\dot{a}^{3}\dot{\phi}+3ka\left( N-\frac{8}{N}\frac{\dot{a%
}}{a}\dot{\phi}\right) +Na^{3}V\left( \phi\right) ,  \label{cc.08}
\end{equation}

from which it appears quite natural to compute the cosmological field
equations with respect to the dynamical variables $\left\{ N,a,\phi\right\} $%
. Specifically, the variation with respect to the lapse function, $N$, gives the
constraint equation.%
\begin{equation}
3H^{2}-24H^{3}\frac{\dot{\phi}}{N}-\frac{3k}{a^{2}}\left( 1+8H\frac {\dot{%
\phi}}{N}\right) -V\left( \phi\right) =0,  \label{cc.09}
\end{equation}
while varying with respect to the scale factor, $a$, and to the scalar
field, $\phi$, provides the following second-order (for $a(t)$ and $\phi$)
differential equations,

\begin{subequations}
\begin{align}
&\frac{1}{N}\dot{H}+H^{2}+\frac{V_{,\phi}}{24\left( H^{2}+\frac{k}{a^{2}}%
\right)}=0,  \label{cc.10} \\
&\frac{12\ddot{\phi }}{N^2}\left( \frac{k}{a^{2}}+H^{2}\right) ^{2}- 3\left( 
\frac{k}{a^{2}}\right)^{2}+\left( 3H^{4}+\frac{V_{,\phi}}{8}\right)-12\left( 
\frac{k}{a^{2}}+H^{2}\right)^{2}\left( \frac{3}{N}H+\frac{\dot{N}}{N^{3}}%
\right)-\frac{1}{8}\frac{\dot{\phi}}{N}HV_{,\phi}=0.  \label{cc.11}
\end{align}
Bearing these results in mind, we can accordingly compute the cosmological
dynamics.

\section{Cosmological dynamics}\label{sezione3}

We here work out a detailed phase-space analysis based on Eqs. (\ref%
{cc.09}), (\ref{cc.10}) and (\ref{cc.11}).

Enabling the Hubble rate to space over positive and negative regions, we can
introduce the dimensionless variables \cite%
{Paliathanasis:2024gwp,Millano:2024vju} 
\end{subequations}
\begin{subequations}
\begin{align}
& x=8\frac{\dot{\phi}}{1+H^{2}}, \\
& y=\frac{V\left( \phi \right) }{1+H^{2}}, \\
& \eta =\frac{H}{\sqrt{\left( 1+H^{2}\right) }}, \\
& \omega _{k}=\frac{k}{a^{2}\left( 1+H^{2}\right) }, \\
& \lambda =\frac{V_{,\phi }}{V}, \\
& d\tau =(1+H^2)dt,
\end{align}

which differ from the $H$-normalization approach \cite{Copeland:2006wr}
since the assumption made through these parameters allows us to pass through 
$H=0$, i.e., to let $H$ change sign, as stated above.

Selecting $\tau$ as the \emph{new independent variable}, the field equations
turn into 
\end{subequations}
\begin{align}
\left( \eta^{4}+\left( 1-\eta^{2}\right) \omega_{k}\right) ^{2}\frac {dx}{%
d\tau} & =-\frac{y\eta^{2}}{24}\left( 1-\eta^{2}\right) \left(
\eta^{2}\left( 2\lambda-\eta\left( 2\left( \lambda-12\right) \eta+\lambda
x\left( 2+\eta^{2}\right) \right) \right) \right)  \notag \\
& -\frac{y\eta^{2}}{24}\left( 1-\eta^{2}\right) ^{2}\left( \lambda
x\eta-24\right) \omega_{k}^{2}+\omega_{k}^{4}\left( 1-\eta^{2}\right)
^{2}\left( \eta^{2}\left( \eta x-1\right) -1\right)  \notag \\
& +\left( \eta^{7}\left( \eta\left( 1+\eta^{2}\left( \eta x-1\right) \right)
-2x\left( 1-\eta^{2}\right) \omega_{k}^{2}\right) \right) ,  \label{cc.12}
\end{align}%
\begin{equation}
\frac{dy}{d\tau}=-\frac{1}{48}y\sqrt{1-\eta^{2}}\left( 3\lambda x+2\eta
^{3}\left( 24+\frac{\lambda y\left( 1-\eta^{2}\right) }{\eta^{4}-\left(
1-\eta^{2}\right) \omega_{k}}\right) \right) ,  \label{cc.13}
\end{equation}%
\begin{equation}
\frac{d\eta}{d\tau}=\frac{y}{24\eta}\left( 3\lambda
x\eta+48\eta^{4}+2\lambda y\left( 1-\eta^{2}\right) \right) ,  \label{cc.14}
\end{equation}%
\begin{equation}
\frac{d\omega_{k}}{d\tau}=\frac{\left( 1-\eta_{k}\right) ^{2}}{12\eta}%
\omega_{k}^{2}\left( 24+\frac{\lambda y\left( 1-\eta^{2}\right) }{\eta
^{4}-\left( 1-\eta^{2}\right) \omega_{k}}\right),  \label{cc.15}
\end{equation}
and%
\begin{equation}
\frac{d\lambda}{d\tau}=\frac{\lambda}{8}x\left( \Gamma\left( \lambda\right)
-1\right) ,  \label{cc.15a}
\end{equation}
where $\Gamma\left( \lambda\right) =\frac{V_{,\phi\phi}V}{\left( V_{,\phi
}\right) ^{2}}$.

Also, from Eq. (\ref{cc.09}) the algebraic constraint becomes
\begin{equation}
\frac{1}{3}y+\omega_{k}-\eta^{2}+\frac{\eta}{1-\eta^{2}}x\left( \eta
^{2}+\omega_{k}\right) =0.  \label{cc.16}
\end{equation}

By construction, $\eta$ is limited as $\eta^{2}\leq1$, although the other
variables are not compactified. Indeed, for $\eta^{2}=1$, the solution
describes a scaling universe, while for $\eta^{2}<1$, the solution describes
a de Sitter solution. When $\eta>0$ the dynamical system
describes an expanding universe $\left( H>0\right)$, while for $\eta<0$, the
dynamical system describes a collapsing universe $\left( H<0\right) $.

In terms of the dimensionless variables the deceleration parameter $q=-1-%
\frac{\dot{H}}{H^{2}}$ \cite{2016IJGMM..1330002D,2012PhRvD..86l3516A,
2014PhRvD..90d3531A, 2017PDU....17...25A, 2024A&A...690A..40L,
2020A&A...641A.174L} is expressed as 
\begin{equation}
q\left( x,y,\eta,\omega_{k}\right) =\frac{\lambda}{24}\frac{1-\eta^{2}}{%
\eta^{2}}\frac{y}{\eta^{2}+\omega_{k}}.  \label{cc.17}
\end{equation}

Immediately we notice that,

\begin{itemize}
\item[-] The field equations form a five-dimensional dynamical system of
first-order differential equations. By using the constraint provided in Eq. (%
\ref{cc.16}), the dimension of the dynamical system is reduced by one.

\item[-] Moreover, for the exponential potential $V\left( \phi\right)
=V_{0}e^{\lambda_{0}\phi}$, it follows that $\Gamma\left( \lambda\right)
-1=0 $, thus, parameter $\lambda$ is always a constant, i.e. $%
\lambda=\lambda_{0}. $

\item[-] Hence, in order to reduce further the dimension of the dynamical
system we consider the exponential potential.

Indeed, for the exponential potential $V\left( \phi\right) $, the
corresponding $f\left( G\right) $ function is derived%
\begin{equation}
f\left( G\right) =\frac{G}{\lambda_{0}}\left[ \ln\left( -\frac{G}{%
\lambda_{0}V_{0}}\right) -1\right] .  \label{cc.18}
\end{equation}
\end{itemize}

\section{Consequences on the observable Universe}\label{sezione4}

We can now wonder whether our scenario can be matched with the observable universe. This would be important in order to check
the goodness of our approach and to infer limits about the initial 
Lagrangian, proposed in Eq. \eqref{cc.03}. To do so, we focus below on spatially-flat and non-flat universe
and find the possible attractors and stability at infinity.

\subsection{Spatially-flat universe}

Consider now a spatially flat FLRW universe, $k=0$, rewriting $y$ from Eq. (%
\ref{cc.16}), the two-dimensional system yields, $\left\{ x,\eta\right\} $,
that is,%
\begin{equation}
\frac{dx}{d\tau}=1-\eta^{2}\left( 1-x\eta\right) -\frac{\left( \eta\left(
x+\eta\right) -1\right) \left( \lambda x\eta\left( 2+\eta^{2}\right)
-2\lambda+2\eta^{2}\left( \lambda-12\right) \right) }{8\eta^{2}},
\label{cc.19}
\end{equation}%
\begin{equation}
\frac{d\eta}{d\tau}=-\frac{1}{24}\eta^{2}\left( 1-\eta^{2}\right) \left( 24+%
\frac{\lambda\eta^{2}\left( 1-\eta^{2}\right) \left( 1-x\eta+\eta
^{2}\right) }{1-\eta^{2}}\right) .  \label{cc.20}
\end{equation}

The stationary/equilibrium points~$A=\left( x\left( A\right) ,\eta\left(
A\right) \right) $ of the latter two-dimensional system are%
\begin{align*}
A_{1} & =\left( 0,1\right) ,~A_{2}=\left( 0,-1\right) ,~ \\
A_{3} & =\left( \frac{32}{3}\frac{1}{\lambda},1\right) ~,~A_{4}=\left( -%
\frac{32}{3}\frac{1}{\lambda},-1\right) , \\
A_{5} & =\left( 0,\sqrt{\frac{\lambda}{\lambda-8}}\right) ,~A_{6}=\left( 0,-%
\sqrt{\frac{\lambda}{\lambda-8}}\right) .
\end{align*}

The equilibrium points with odd numbers describe expanding universes, while
the equilibrium points with even numbers describe collapse. Points $A_{5}$, $%
A_{6}$ describe de Sitter solutions, while the remaining points describe
scaling exact solutions.

In particular, the equilibrium points $A_{1}$,~$A_{2}$ describe asymptotic
solutions with deceleration parameters $q\left( A_{1}\right) =0,~$ $q\left(
A_{2}\right) =0$. The Gauss-Bonnet term contributes in the cosmic fluid, and 
$y\left( A_{1}\right) =3,~y\left( A_{2}\right) =3$. As far as the stability
of the points is concerned, we determine that the eigenvalues of the
linearized system around point $A_{1}$ are $\left\{ 4,2\right\} $, and for
point $A_{2}$ they are $\left\{ -4,-2\right\} $. Thus point $A_{1}$ is
always a source and point $A_{2}$ is always an attractor.

Equilibrium points $A_{3}$ and $A_{4}$ describe scaling solutions which can
describe cosmic acceleration, that is, $q\left( A_{3}\right) =-\frac{4}{3}$
and $q\left( A_{4}\right) =-\frac{4}{3}$. At these points parameter $y$
reaches infinity. The eigenvalues for point $A_{3}$ are $\left\{ -4,-\frac {2%
}{3}\right\} $, while for point $A_{4}$ are $\left\{ 4,\frac{2}{3}%
\right\} $. Therefore, point $A_{3}$ is always an attractor and point $A_{4}$
is always a source point.

Furthermore, equilibrium points $A_{5}$ and $A_{6}$ describe de Sitter
solutions with $q\left( A_{5}\right) =-1$, and $q\left( A_{6}\right) =-1$.
The points are real for $\lambda<0.$ At these points there is a nonzero
contribution of the scalar field potential, that is, $y=\frac{3\lambda }{%
\lambda-8}$. The eigenvalues of the linearized systems around these points
are $\left\{ -\frac{3+\sqrt{17}}{2}\sqrt{\frac{\lambda}{\lambda-8}},-\frac{3-%
\sqrt{17}}{2}\sqrt{\frac{\lambda}{\lambda-8}}\right\} $, from where we
conclude that the equilibrium points are always saddle points.

\subsubsection{Analysis at infinity}

Above, we computed the equilibrium points for the dynamical system, reported
in Eqs. (\ref{cc.19}) and (\ref{cc.20}), in the finite regime.

Nevertheless, clarifying the asymptotic regime appears quite interesting, as
the variable $x$ can acquire values at infinity.

Accordingly, we thus introduce the new compactified variable, $X$, so that $%
x=\frac{X}{\sqrt{1-X^{2}}}$ and study the existence of equilibrium points at
infinity, where $X^{2}=1$.

Indeed, there are four points, $\hat{A}=\left( X\left( \hat{A}\right)
,\eta\left( \hat{A}\right) \right) ,$ with coordinates%
\begin{equation*}
\hat{A}_{1}^{\pm}=\left( \pm1,1\right) ~,~\hat{A}_{2}^{\pm}=\left(
\pm1,-1\right) .
\end{equation*}
At these points we calculate the deceleration parameter $q\left( \hat{A}%
_{1}^{\pm}\right) =-sign\left( \lambda X\left( \hat{A}_{1}^{\pm}\right)
\right) \infty$ and $q\left( \hat{A}_{2}^{\pm}\right) =sign\left( \lambda
X\left( \hat{A}_{2}^{\pm}\right) \right) \infty$. Points $\hat{A}_{1}^{\pm }$
describe an expanding universes, but for $sign\left( \lambda X\left( \hat {A}%
_{1}^{\pm}\right) \right) <0,$ we derive that $q\left( \hat{A}_{1}^{\pm
}\right) >0$, that means these stationary points correspond to asymptotic
solutions, asymptotically reaching the Minkowski space.

The same holds for the points exhibiting $sign\left( \lambda X\left( \hat{A}%
_{2}^{\pm}\right) \right) <0$.

Consequently, the existence of these stationary points is essential in order
the cosmological evolution to cross from expansion to collapse and vice
versa. Recall that $\eta=0$, gives a singularity at the dynamical system, so
the transition is happening through the points at infinity.

\begin{figure}[ptbh]
\centering\includegraphics[width=1\textwidth]{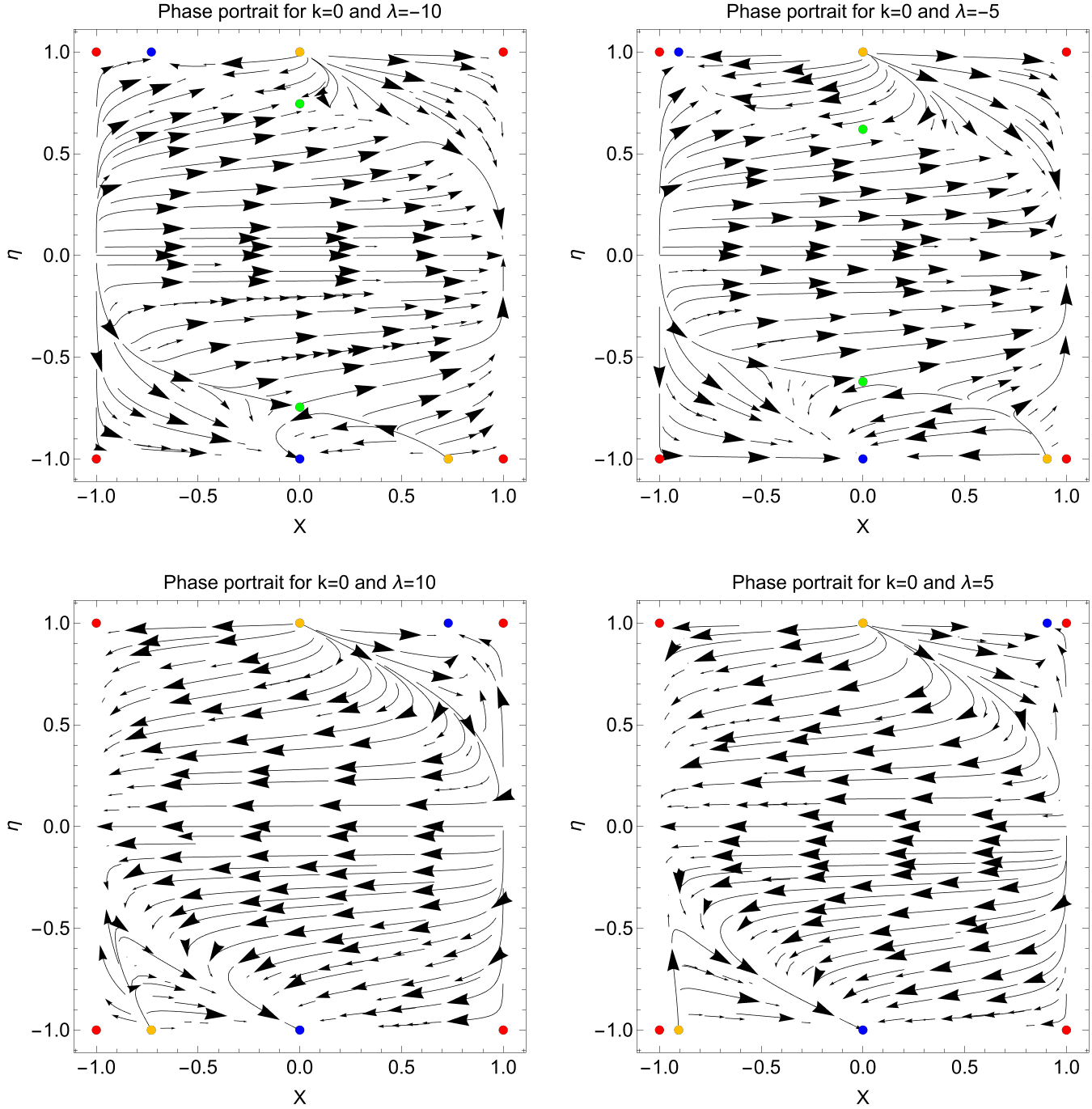}
\caption{Phase-space portrait for the dynamical system (\protect\ref{cc.19}%
), (\protect\ref{cc.20}) in the compactified variables $X$ and $\protect\eta$%
. Equilibrium points at the infinity are marked with red, the source points $%
A_{1}$ and $A_{4}$ are marked with orange, the saddle points $A_{5}$ and $%
A_{6}$ with green, and the attractors $A_{2}$ and $A_{5}$ are marked with
blue. We observe that the main of the initial conditions lead to a collapsed
universe, while there exist only a small area, less than then one quarter,
of the phase-space, where the trajectories lead to a future attractor which
describes cosmic acceleration.}
\label{fig1}
\end{figure}

This prescription is explicitly reported in Fig. \ref{fig1}. Last but not
least, the stationary points at infinity always correspond to unstable
solutions.

From the phase-space portraits displayed in Fig. \ref{fig1}, it appears clear that the
most of underlying initial conditions leads to a collapsed universe, whereas
only a small fraction of the entire phase-space, less than one quarter,
corresponds to trajectories with initial conditions leading to a future
attractor and, then, describing cosmic acceleration.

A summary of the above outcomes is summarized in Table \ref{tab11}.

\begin{table}[tbp] \centering%
\caption{Equilibrium Points for the spatially flat FLRW universe and
exponential potential.}%
\begin{tabular}{ccccc}
\hline\hline
\textbf{Point} & $\left( \mathbf{x,\eta}\right) $ & \textbf{Existence} & 
\textbf{$q$} & \textbf{Stable?} \\ \hline
$A_{1}$ & $\left( 0,1\right) $ & Always & $0$ & False \\ 
$A_{2}$ & $\left( 0,-1\right) $ & Always & $0$ & Attractor \\ 
$A_{3}$ & $\left( \frac{32}{3}\frac{1}{\lambda},1\right) $ & $\lambda\neq0$
& $-\frac{4}{3}$ & Attractor \\ 
$A_{4}$ & $\left( -\frac{32}{3}\frac{1}{\lambda},-1\right) $ & $\lambda
\neq0 $ & $-\frac{4}{3}$ & False \\ 
$A_{5}$ & $\left( 0,\sqrt{\frac{\lambda}{\lambda-8}}\right) $ & $\lambda<0$
& $-1$ & False \\ 
$A_{6}$ & $\left( 0,-\sqrt{\frac{\lambda}{\lambda-8}}\right) $ & $\lambda<0$
& $-1$ & False \\ 
$\hat{A}_{1}^{\pm}$ & $\left( \pm\infty,1\right) $ & Always & $\mp
sign\left( \lambda\right) \infty$ & False \\ 
$\hat{A}_{2}^{\pm}$ & $\left( \pm\infty,-1\right) $ & Always & $\pm
sign\left( \lambda\right) \infty$ & False \\ \hline\hline
\end{tabular}
\label{tab11}%
\end{table}%

\subsection{Non-zero spatial curvature}

Let us now assume now that the spatial curvature does not vanish.
Accordingly, the variable $\omega_{k}$ furnishes a non-trivial dynamics.

We determine the equilibrium points $B=\left( x\left( B\right) ,\eta\left(
B\right) ,\omega_{k}\left( B\right) \right) $ for the dynamical system in
Eqs. (\ref{cc.12}), (\ref{cc.14}) and (\ref{cc.15}), reading 
\begin{align*}
B_{1} & =\left( 0,1,\omega_{k}\right) ,~B_{2}=\left( 0,-1,\omega _{k}\right)
, \\
B_{3} & =\left( \frac{8}{3\lambda}\frac{4+3\omega_{k}}{1+3\omega_{k}}%
,1,\omega_{k}\right) , \\
B_{4} & =\left( -\frac{8}{3\lambda}\frac{4+3\omega_{k}}{1+3\omega_{k}}%
,-1,\omega_{k}\right) , \\
B_{5} & =\left( 0,\sqrt{\frac{\lambda}{\lambda-8}},0\right) ,~B_{6}=\left(
0,-\sqrt{\frac{\lambda}{\lambda-8}},0\right),
\end{align*}
where at points $B_{1},~B_{2},~B_{3}$ and $B_{4}$,~$\omega_{k}$ is undefined
since these do not represent just points, but surfaces defined on the
parametric space with varying $\omega _{k}$. Hence, the solution
trajectories lie on these surfaces and move along $\omega _{k}$.

The equilibrium points $B_{1},~B_{2}$ describe cosmic expansion and collapse
respectively. The corresponding deceleration parameters are $q\left(
B_{1}\right) =0$ and $q\left( B_{2}\right) =0$. The eigenvalues, computed
near $B_{1}$, are $\left\{ 2,4+3\omega_{k},0\right\} $, while near $B_{2}$,
they are $\left\{ -2,-4-3\omega_{k},0\right\} $. The equilibrium point $%
B_{1} $ is always unstable, differently from $B_{2}$, for which there can be
a stable submanifold.

To check this, we expand up to the second-order approximation, seeking the $%
\omega_k$ perturbations, and find the exact solution for the evolution of
the perturbations $\delta x\simeq e^{-2t}+\alpha_{0}e^{-\left(
4+3\omega_{k}\right) t}$,~$\delta\eta\simeq e^{-2t}$ and $\delta\omega
_{k}\simeq e^{-4t}+\alpha_{1}e^{-3\left( 2+\omega_{k}\right) t}$ from which
we conclude that $B_{2}$ turns out to be stable.

Furthermore, the equilibrium points $B_{3}$ and $B_{4}$ are defined for $%
\omega_{k}\neq-\frac{1}{3}$ and appear the generalizations of $A_{3}$ and $%
A_{4}$, respectively. The deceleration parameters are $q\left( B_{3}\right)
=-1-\frac{1}{3\left( 1+\omega_{k}\right) }$ and $q\left( B_{4}\right) =-1-%
\frac{1}{3\left( 1+\omega_{k}\right) }$. Thus, here the cosmic acceleration is recovered either for $\omega_{k}<-%
\frac{4}{3}$ or $\omega_{k}>-1$.

The eigenvalues of the linearized system around the point $B_{3}$ are $%
\left\{ -4-3\omega_{k},-\frac{2}{3}\left( 1+3\omega_{k}\right) ,0\right\} ,$
while at the point $B_{4}$ we find $\left\{ 4+3\omega_{k},\frac{2}{3}\left(
1+3\omega_{k}\right) ,0\right\} $.

In analogy to what found before, we study the evolution of the perturbations
up to the second-order and it follows that $B_{3}$ appears as an attractor
for $\omega_{k}>-\frac{1}{3}$, while $B_{4}$ is an attractor with $%
\omega_{k}<-\frac{4}{3}$.

Finally, the stationary points $B_{5},$ $B_{6}$ describe a de Sitter
solution of the spatially flat universe of the points $A_{5}$ and $A_{6}$.

Last but not least we found that $B_{5}$, $B_{6}$ share the two eigenvalues
with points $A_{5}$ and $A_{6}$, from which we infer that they are always
saddle points.

\subsubsection{Analysis at infinity}

We employ the same compactified variable $X$ as before, and we find that the
stationary points at the infinity are 
\begin{equation*}
\hat{B}_{1}^{\pm}=\left( \pm1,1,\omega_{k}\right) ~,~\hat{B}_{2}^{\pm
}=\left( \pm1,-1,\omega_{k}\right) .
\end{equation*}
They provide the same physical properties as in the flat case, while
remarkably the stationary points always describe unstable solutions.

If we consider that $\left\vert \omega_{k}\right\vert $ is small, then the
future attractors are the points $B_{2}$ and $B_{3}$.

These points may exhibit nonzero curvature. However, since the phase-space
trajectories pass through a de Sitter expansion, in which $\eta ^{2}<1$,
then the parameter $\left\vert \omega _{k}\right\vert $ tends to decrease,
when the attractor is point $B_{3}$. This may be reinterpreted as a solution
of the flatness problem. In order to demonstrate this, consider that we
start from initial conditions of a nonzero curvature FLRW space. Then the
trajectories will move along the surfaces we described before. Parameter $%
0<\eta <1$, gives necessary a de Sitter expansion, since~for $\eta _{0}$%
{, }$H=\frac{\eta _{0}}{\sqrt{1-\eta _{0}^{2}}}$. Then $\left\vert
\omega _{k}\right\vert =\left\vert \frac{k}{a^{2}\left( 1+H^{2}\right) }%
\right\vert $, $\downarrow $, because $a\uparrow $, then $\omega
_{k}\rightarrow 0$, asymptotically. .

Otherwise in the collapse, i.e., as the point $B_{2}$ is the attractor
parameter, then $\left\vert \omega_{k}\right\vert $ tends to increase.
However, information toward the original sign of curvature of the dynamical
system remains unaltered at the attractors.

The phase-space behavior is thus analogous to that displayed in Fig. \ref%
{fig1}, where it turns out that the most initial conditions result into a
gravitational collapse.

{In term of dynamics the curvature component can be seen equivalently in the
same form of a spatially flat geometry induced with a indeal gas with
constant equation of state $p_{m}=-\frac{1}{3}\rho _{m.}$. By using this
analogue, we can compare our results with the matter dominated investigated
in \cite{Goheer:2009qh}. Nevertheless this is an asymptotic solution, near a
specific value of the Gauss-Bonnet scalar because $x=0$, that is, the function (\ref{cc.18} ) is described by a polynomial similar to that determined in \cite{Goheer:2009qh}. }

\section{Fixing the scalar field potential}

\label{sezione5}

The use of exponential scalar field, as above reported, is quite usual in
the literature \cite%
{Paliathanasis:2024gwp,Millano:2024vju,Copeland:2006wr,Carloni:2023egi}. The
advantage of using such a potential is to reduce the complexity of variables
under exam. However, more appropriate scalar field potentials may be even
associated with dark energy scenarios, see e.g. \cite%
{2024arXiv241010935C,Lazkoz:2007mx,Paliathanasis:2024jxo,Papagiannopoulos:2016dqw}
and, then, turn out to be relevant to check the validity of our background
theory. Below, we employ a class of power law potentials, with unfixed
exponents, to explore the impact in our stability treatment.

\subsection{The role of power-law potential}

All our treatment made use of the exponential potential, for which Eq. (\ref%
{cc.15a}) turns out to be trivially satisfied, yielding a constant $\lambda$.

For completeness, it appears essential to explore alternative potentials, in
order to evaluate the impact to dynamical variables, within the usual
context of a spatially flat FLRW geometry.

The simplest approach is based on ensuring the validity of a power-law
potential,

\begin{equation}
V\left( \phi \right) =V_{0}\phi ^{n},  \label{powerlaw}
\end{equation}%
from where it follows $\lambda =\frac{n}{\phi }$ and $\Gamma \left( \lambda
\right) =1-\frac{1}{n}$, obtained by definition, since $\Gamma \left( \phi
\right) =\frac{n\left( n-1\right) \phi ^{n-2}\phi ^{n}}{n^{2}\left( \phi
^{n-1}\right) ^{2}}=\frac{n-1}{n}$.

In view of Eq. \eqref{powerlaw}, Eq. (\ref{cc.15a}) furnishes%
\begin{equation}
\frac{d\lambda}{d\tau}=\frac{1}{8n}\lambda x.  \label{cc.15b}
\end{equation}
Remarkably, for this potential we calculate the power-law $f\left( G\right) $
model, that approximately reads $f\left( G\right) \simeq G^{\frac{n}{n-1}}$.

The equilibrium points $C=\left( x\left( C\right) ,\eta\left( C\right)
,\lambda\left( C\right) \right) $ of the dynamical system, namely Eqs. (\ref%
{cc.19}), (\ref{cc.20}) and (\ref{cc.15b}), are%
\begin{align*}
C_{1} & =\left( A_{1},\lambda\right) ,~C_{2}=\left( A_{2},\lambda\right) ,~
\\
C_{3} & =\left( A_{5},\lambda\right) ,~C_{4}=\left( C_{6},\lambda\right),
\end{align*}
where the parameter $\lambda$ above, for all these points, appears
arbitrary. Phrasing it differently, the points actually describe a family of
solutions for different values of $\lambda$. In the stationary points $%
\lambda $ is varying without affective the solution trajectories, in a
similar way as before for the $\omega _{k}$. When $\phi \rightarrow \infty $, for $n>0$, or $\phi \rightarrow 0$, for $n<0$, the potential term
dominates, and there is not any contribution from the potential term, i.e. $%
x=0$, that exactly is happening here. The value of the parameter $\lambda$
is defined by the initial conditions.

We remark that all the equilibrium points have $x=0$. 

The physical properties of the asymptotic solutions are described by the corresponding
four points for the exponential potential. It is clear that the scaling
solutions described by points $A_{3}$,~$A_{4}$ do not exist in this
consideration. 

Point $C_{1}$ is found to always be a source, while point $C_{2}$ is an attractor.

For the equilibrium point $C_{4}$ we calculate the eigenvalues,

\begin{equation}
\left\{ 0,\frac{3\sqrt{n\lambda}-\sqrt{17n\lambda+8}}{2\sqrt{n\left( \lambda
-8\right) }},\frac{3\sqrt{n\lambda}+\sqrt{17n\lambda+8}}{2\sqrt{n\left(
\lambda-8\right) }}\right\},
\end{equation}
from where conclude that the point describes an unstable solution.
Specifically it is a saddle point for $n>-\frac{1}{\lambda }$, provided that
the existence condition requires $\lambda<0$.

Finally, point $C_{3}$ describes a de Sitter expanding universe and the
corresponding eigenvalues are

\begin{equation}
\left\{ 0,\frac{-3\sqrt{n\lambda}-\sqrt{17n\lambda+8}}{2\sqrt{n\left(
\lambda-8\right) }},\frac {-3\sqrt{n\lambda}+\sqrt{17n\lambda+8}}{2\sqrt{%
n\left( \lambda-8\right) }}\right\}.
\end{equation}

Thus, the two eigenvalues are negative when $\lambda<0$ and $0<n<-\frac{8}{%
17\lambda},$ or$~-\frac{8}{17\lambda}<n<-\frac{1}{\lambda}$. From the
analysis of the second-order expansion within the linearized system, we
conclude that the point has a stable submanifold. Specifically, it was found
that for initial conditions where $\lambda<0$, and $n>0$, the surface of
points $C_{3}$ describes stable de Sitter solutions.

\subsection{Asymptotic analysis at Infinity}

Equilibrium points at infinity exist only as $\lambda=0$ and their physical
properties are those described by $\hat{A}^{\pm}$.

We end up that none of the equilibrium point at infinity is an attractor.

\begin{figure}[ptbh]
\centering\includegraphics[width=1\textwidth]{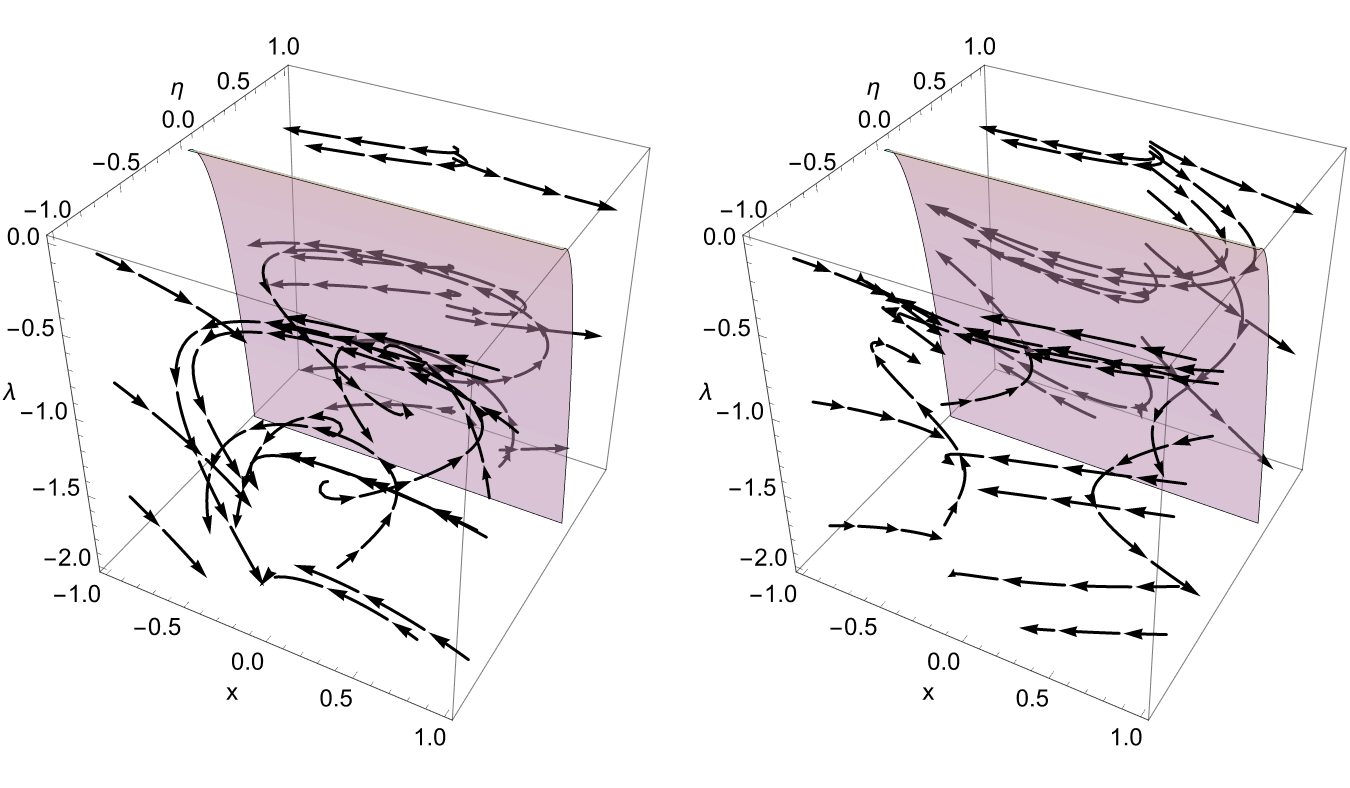}
\caption{Phase-space portrait for the dynamical system (\protect\ref{cc.19}%
), (\protect\ref{cc.20}) and (\protect\ref{cc.15b}) in the variables $x,%
\protect\eta~$and $\protect\lambda$ where the family of points $C_{3}$ is
marked. Left Fig. is for $n=\frac{1}{5}$ and right Fig. for \thinspace$n=-%
\frac{1}{5}$ . It is easy to see that for $n=\frac{1}{5}$, the trajectories
lie on the stable surface $C_{3}$, and the de Sitter solution is an
attractor. }
\label{fig2}
\end{figure}

In Fig.\ref{fig2}, we present three-dimensional phase-portraits for two
values of parameter $n$ and remark that, for $n>0,$ there exists a region
where the trajectories reach on the stable surface of de Sitter expanding
solutions, described by the family of points $C_{3}$.

\subsection{Discussion for a general potential}

{Consider now the generic nonlinear potential }$V\left( \phi \right) $%
{, that is, a generic nonlinear function }$f\left( G\right) ${%
. Then the equilibrium points solve the algebraic equation (\ref{cc.15a}),
that is, }%
\begin{equation*}
\frac{\lambda }{8}x\left( \Gamma \left( \lambda \right) -1\right) =0
\end{equation*}%
{that is, }$x=0${\ or }$\lambda =\lambda _{0}${\ such
that }$\lambda _{0}\left( \Gamma \left( \lambda _{0}\right) -1\right) =0$%
{, where }$\Gamma \left( \lambda \right) ${\ is constructed by
the potential.}

{When }$x=0${, it follows from before that }$\lambda ${%
\ does not constrained by the rest equations of the dynamical system, that
is, the admitted equilibrium points are those described by the power-law
potential.}

{On the other hand, for }$\lambda =\lambda _{0}${, then from
the rest of the equations variables }$x,\eta ${\ are constrained and
provide families of equilibrium points as given by the exponential
potential, for each value of }$\lambda _{0}${. }

{Therefore, the definition of an arbitrary potential function will
not lead to the derivation of new families of solutions. Hence, the study of
the exponential and the power-law potentials is sufficient to construct
all the possible asymptotic solutions. }

{However, what is important to mention is that the stability
properties of the equilibrium points will not remain the same. Because the
function }$\Gamma \left( \lambda \right) ${ and, its derivative play
an important role in the linearization of the dynamical system around the
equilibrium points. }

\section{Final remarks}

\label{sezione6}

In this work, we explored a field-equivalent representation of $f(G)$%
-theories of gravity, with the aim of investigating the phase-space dynamics
and the corresponding stability. To do so, we specifically considered a FLRW
universe with flat topology and introduced a Lagrange multiplier within the
modified Lagrangian density, \emph{de facto} reformulating the action of $%
f(G)$-gravity into an equivalent form of Einstein-Gauss-Bonnet scalar field
cosmology, that does not exhibit an explicit scalar field kinetic term.

The scalar field potential is directly related to the function $%
f(G)$ and its derivatives. Further, within this scalar field formulation, we
computed the cosmological field equations and introduced a set of
dimensionless variables, quite different from those used in standard
approach, i.e., carrying out a global dynamical analysis by enabling the
Hubble rate to change its sign.

In view of the fact that the field equations constituted a complicated
constrained dynamical system, by utilizing the constraint equation, we
reduced the system's dimensionality by one, reducing to four. In addition,
we showed that, using an exponential potential, the dynamical system is
further reduced by another dimensionality. Afterwards, we identified the
equilibrium points and found two attractors, both independent from the
spatial curvature.

Within our analysis, one equilibrium point described an expanding,
accelerating universe, not behaving as a de Sitter solution, while another
one corresponded to a collapsing spacetime.

More precisely, even though a de Sitter solution exists, it manifests as a
saddle point. Nevertheless, the phase-space analysis indicated that most
initial conditions lead to a collapsed universe, pointing out that our
scenario is likely unphysical.

Indeed, in the power-law $f(G)$-theory, the scaling solution appears absent,
and the de Sitter solution is a future attractor for positive
values of the power parameter and for specific set of initial conditions. {That means that while the $f(G)$-theory can provide an explanation to the cosmic acceleration, the majority of the trajectories of the solutions lead to attractors which can not describe the present cosmic acceleration. Nevertheless, unstable equlibrium points which describe cosmic acceleration can be related to the early-time inflation.}

Future works will try to extend our treatment to more physical contexts in
which the topological terms may imply more stable regions, whose physical
sense can be more defined. In particular, by adding further external fields,
such as sources to the energy momentum-tensor or trying to include
additional constraints. The impact of perturbations will be also
investigated, in order to fix more carefully the free parameters of the
theory, as well as possible additional potentials, based on precise physical
meanings, related to dark energy sources \cite{2024arXiv241010935C} or
inflationary domains.

\begin{acknowledgments}
GL and AP are grateful for the support of Vicerrector\'{\i}a de Investigaci%
\'{o}n y Desarrollo Tecnol\'{o}gico (Vridt) at Universidad Cat\'{o}lica del
Norte through N\'{u}cleo de Investigaci\'{o}n Geometr\'{\i}a Diferencial y
Aplicaciones, Resoluci\'{o}n Vridt No - 096/2022 and Resoluci\'{o}n Vridt No
- 098/2022. GL and AP were economically supported by the Proyecto Fondecyt
Regular 2024, Folio 1240514, Etapa 2024. AP expresses his gratitude to the
University of Camerino and to OL for invitation and hospitality provided
during the time in which this work has been written. OL acknowledges
financial support from the Fondazione ICSC, Spoke 3 Astrophysics and Cosmos
Observations. National Recovery and Resilience Plan (Piano Nazionale di
Ripresa e Resilienza, PNRR) Project ID CN$\_$00000013 "Italian Research
Center on High-Performance Computing, Big Data and Quantum Computing" funded
by MUR Missione 4 Componente 2 Investimento 1.4: Potenziamento strutture di
ricerca e creazione di "campioni nazionali di R$\&$S (M4C2-19 )" - Next
Generation EU (NGEU) GRAB-IT Project, PNRR Cascade Funding Call, Spoke 3,
INAF Italian National Institute for Astrophysics, Project code CN00000013,
Project Code (CUP): C53C22000350006, cost center STI442016.
\end{acknowledgments}

\bibliographystyle{unsrt}
\bibliography{bibliography}

\end{document}